\documentclass[preprint,12pt]{elsarticle}

\usepackage{amssymb} 
\usepackage{bm}
\usepackage[version=3]{mhchem} 
\usepackage{amsthm} 
\usepackage{caption} 

\usepackage{booktabs}
\usepackage{xcolor}
\usepackage[normalem]{ulem}
\usepackage{here}

\journal{Computational Materials Science}

\begin{document}

\begin{frontmatter}



\title{Crystal structure prediction with machine learning-based element substitution}


\date{}

\author[inst1]{Minoru Kusaba\corref{cor1}}
\ead{kusaba@ism.ac.jp}

\author[inst2]{Chang Liu}

\author[inst1,inst2,inst3]{Ryo Yoshida\corref{cor1}}
\ead{yoshidar@ism.ac.jp}

\affiliation[inst1]{organization={The Graduate University for Advanced Studies},
            addressline={SOKENDAI}, 
            city={Tachikawa},
            postcode={190-8562}, 
            state={Tokyo},
            country={Japan}}
            
\affiliation[inst2]{organization={The Institute of Statistical Mathematics},
            addressline={Research Organization of Information and Systems}, 
            city={Tachikawa},
            postcode={190-8562}, 
            state={Tokyo},
            country={Japan}}

\affiliation[inst3]{organization={National Institute for Materials Science},
            addressline={Research Organization of Information and Systems}, 
            city={Tsukuba},
            postcode={305-0047}, 
            state={Ibaraki},
            country={Japan}}

\cortext[cor1]{Corresponding authors}

 \begin{abstract}
The prediction of energetically stable crystal structures formed by a given chemical composition is a central problem  in solid-state physics. In principle, the crystalline state of assembled atoms can be determined by optimizing the energy surface, which  in turn can be evaluated using first-principles calculations. However, performing the iterative gradient descent on the potential energy surface using first-principles calculations is prohibitively expensive for complex systems, such as those with many atoms per unit cell. Here, we present a unique methodology for crystal structure prediction (CSP) that relies on a machine learning algorithm called metric learning. It is shown that a binary classifier, trained on a large number of already identified crystal structures, can determine the isomorphism of crystal structures formed by two given chemical compositions with an accuracy of approximately 96.4\%. For a given query composition with an unknown crystal structure, the model is used to automatically select from a crystal structure database a set of template crystals with nearly identical stable structures to which element substitution is to be applied.  Apart from the local relaxation calculation of the identified templates, the proposed method does not use ab initio calculations. The potential of this substitution-based CSP is demonstrated for a wide variety of crystal systems.
\end{abstract}


\end{frontmatter}


\section{Introduction}
\label{sec:section1}
Predicting the crystal structure formed by an arbitrary chemical composition remains an unsolved problem in solid-state physics. In principle, stable or metastable structures formed by atomic or molecular assemblies can be found by solving a local optimization problem for a potential energy surface defined on the space of atomic coordinates. A major approach to computational crystal structure prediction (CSP) relies on the repeated calculation of first-principles potential energy surfaces, such as density functional theory (DFT) calculations \cite{hohenberg1964inhomogeneous,kohn1965self}. During these calculations, the random search \cite{pickard2006high,pickard2007structure,pickard2011ab}, simulated annealing \cite{kirkpatrick1983optimization,pannetier1990prediction}, basin-hopping \cite{wales1997global}, minima hopping \cite{goedecker2004minima,amsler2010crystal}, evolutionary algorithm (EA) \cite{oganov2006crystal,oganov2011evolutionary,lyakhov2013new}, particle swarm optimization (PSO) \cite{wang2010crystal,zhang2017computer}, Bayesian optimization (BO) \cite{yamashita2018crystal}, and look ahead based on quadratic approximation (LAQA) \cite{terayama2018fine} have been employed as the optimization methods. More recently, as a promising alternative, machine-learning interatomic potentials, which substantially speed up the optimization by bypassing the time-consuming ab initio calculations, have attracted considerable attention\cite{jacobsen2018fly,podryabinkin2019accelerating}. These methods can be classified into two types: sequential search and batch selection. Sequential search methods, such as EA and PSO, explore the global or local minimum of the potential energy surface. In these methods, a current set of candidate crystalline forms is iteratively modified with a predefined set of genetic manipulations, in which ab initio structural optimization is repeatedly applied to the currently obtained candidates. Batch selection methods such as BO and LAQA utilize surrogate models, learned from a training set of DFT energies and crystal structures, to identify more promising candidates with lower predicted energies from a predefined set of candidate crystals. A reasonable set of initial structures must be created using a crystal structure generator in both the cases. The random symmetric structure generator \cite{lyakhov2013new,zhu2012constrained} and the topology-based structure generator \cite{bushlanov2019topology} have been proposed as the generators. Nonetheless, these methods rely on the iterative use of computationally expensive ab initio energy calculations.

Another type of computational CSP is based on element substitution \cite{hautier2011data,wang2021predicting,https://doi.org/10.48550/arxiv.2111.14049}. Most  crystals synthesized so far have been discovered by considering the element substitution of previously discovered ones. Substitution-based CSP mimics such traditional protocols computationally; it aims to predict the stable crystal structure by replacing elements in an already known template crystal, which possesses high chemical replaceability to the target structure to be predicted. Such substitution-based methods do not require time-consuming potential energy calculations, except in the process of locally optimizing replaced crystals. However, unlike ab initio energy-based methods, template-based methods cannot  predict new crystal structures. Despite such limitations, template-based methods, owing to significantly lower computational costs, are highly useful for the prediction of many crystal structures \cite{hautier2011data}.

Here, we present a powerful CSP method based on machine-learned element substitution. The method relies on a machine-learning algorithm referred to as metric learning \cite{kulis2013metric}. Metric learning is used to automate the selection of template structures from a crystal structure database with high chemical replaceability to the unknown stable structure for a given chemical composition. In metric learning, a binary classifier is constructed to determine whether the crystal structures of two given chemical compositions are identical or non-identical. Crystals with sufficiently high structural similarity are treated as identical, and the labeled  dataset is extracted from the crystal structure database. The prediction accuracy of the trained model exceeds 96.4\%. Solving the inverse problem of the trained classifier by performing a thorough screening over a large number of known crystals, a set of compositions — as well as their crystal structures that are highly replaceable to a given query composition —  can be identified. Then, a template structure is created by assigning the constituent elements in the query composition to the selected template, and a stable crystalline form is obtained by relaxing the created template structure to reach the local minimum energy using DFT calculations. 

The substitution-based method proposed by Wei et al. \cite{https://doi.org/10.48550/arxiv.2111.14049} determines substitution targets based solely on the similarity of elements in pairs of chemical formulas. Therefore, no information on the similarity of crystal structures is used. The substitution-based methods proposed by Hautier et al. \& Wang et al. \cite{hautier2011data,wang2021predicting} statistically estimate the replaceability of two chemical elements based on the observed frequency of their occurrence in two similar crystal structures. As a result, co-occurrence patterns with other elements are completely ignored. Another problem is that, in principle, the model cannot recognize what is dissimilar because the previous methods do not use any data on non-identical structures during model training.
The proposed method improves the prediction accuracy and extends the applicability domain of the model by learning the replaceability of the overall context of chemical compositions, rather than a pair of elements, with training instances from both similar and dissimilar structures. We show that, in estimation, our substitution-based approach can predict stable structures of approximately 50 \% of all crystals discovered so far with high confidence . The code for the CSP method is available at \cite{CSPML}. 

\section{Methods}
\label{sec:section2}
\subsection{Outline}
Let $C_i$ be a chemical composition and $S_i$ be the corresponding stable crystal structure. The chemical composition $C_i$ is
characterized by a descriptor vector $\bm{\phi}(C_i)\in\mathbb{R}^d$ that encodes $d$ features of the constituent elements in $C_i$, as detailed below. For a given pair of chemical compositions $C_i$ and $C_j$, we assign a binary class label $y_{ij}$ which takes the value 1 if the corresponding stable structures $S_i$ and $S_j$ are significantly close, and 0 otherwise.  Here, we construct a model $f$ that predicts the structure similarity label $y_{ij}$ for any given pair of $C_i$ and $C_j$. The model learns via the supervision of known crystals and their compositions in a crystal structure database. The model takes $\bm{\phi}(C_i)$ and $\bm{\phi}(C_j)$ as inputs and outputs a classification probability $f$ representing their structural identity, which serves as a metric for structure similarity or replaceability between $C_i$ and $C_j$. The problem reduces to the task of metric learning \cite{kulis2013metric}.

\begin{figure}[htp]
\centering
\includegraphics[width=12cm]{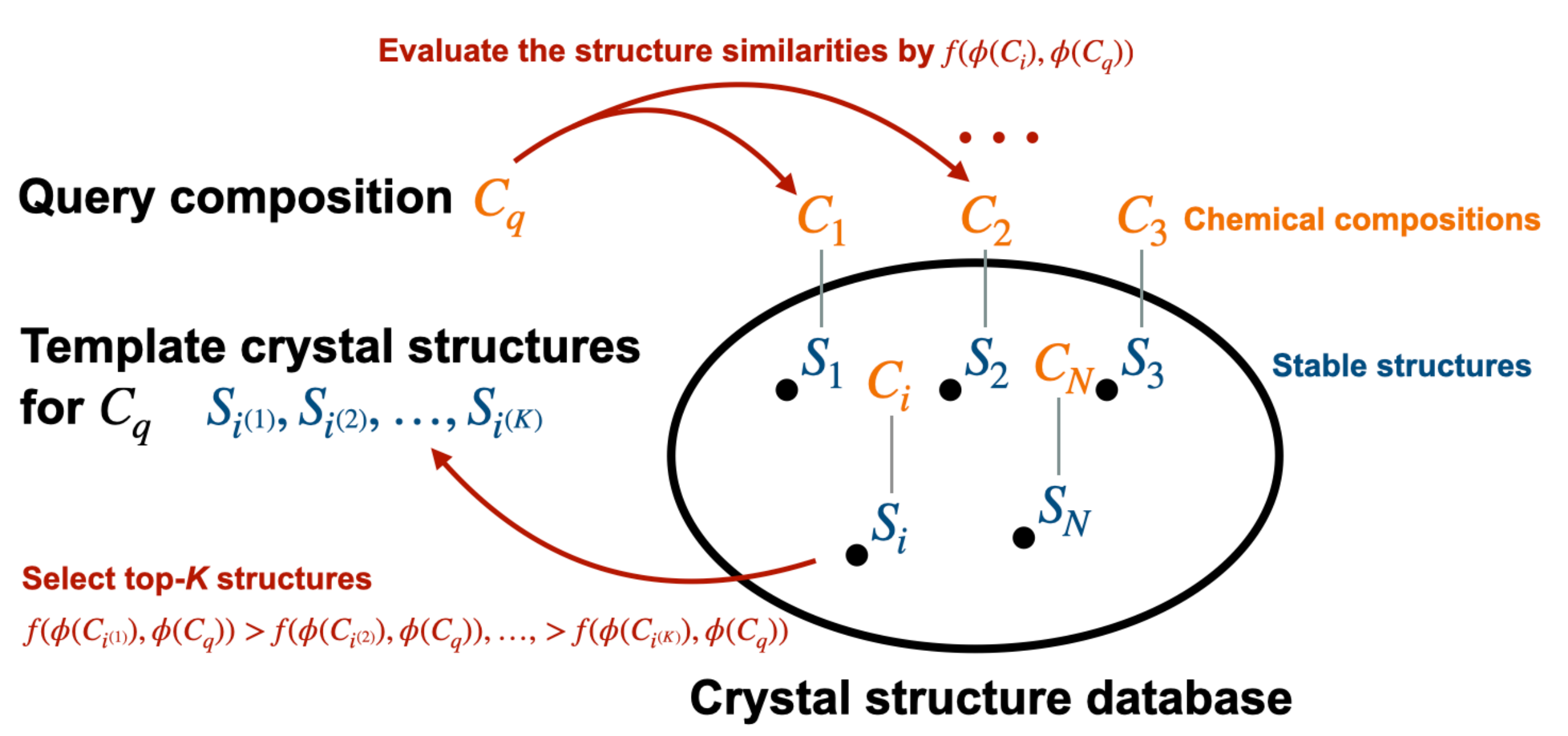}
\caption{Schematic depiction of the substitution-based CSP using metric learning.}
\label{fig:fig1}
\end{figure}

The trained model (metric) $f$ was used for CSP. For a query composition $C_q$, our goal is to predict the stable structure (denoted by $S_q$). Let us assume that the database records $N$ chemical compositions $C_1,\ldots,C_N$ and their stable structures $S_1,\ldots,S_N$. If the database contains crystal structures that are sufficiently close to $S_q$, CSP can be performed by screening out those crystals. We can then evaluate the structure similarity between $S_q$ and $S_i$ $(i=1,\ldots,N)$ by assigning $\bm{\phi}(C_q)$ and $\bm{\phi}(C_i)$ to $f$, and selecting the top-$K$ structures as templates. The element species in $C_q$ are assigned to the atoms in each of the top-$K$ selected structures, which are optimized with the DFT calculation to fine-tune the atomic configuration to decrease the free energy . The workflow is summarized in Fig. \ref{fig:fig1}.

\subsection{Learning to predict structural identity from compositional features}

From the given candidate compounds, we select the crystal structures that are predicted to be similar to $S_q$ using a metric $f$ of structure similarity. The model $f$ is constructed using a metric learning algorithm. A training dataset is prepared by taking compound pairs $\{(C_i,C_j,y_{ij})|i,j = 1,\ldots,M\}$ from the crystal structure database. A structure similarity label is assigned to each $(C_i,C_j)$ by applying a threshold value $\tau = 0.3$ to the crystal structure similarity measure calculated using the local structure order parameters \cite{zimmermann2020local} (see Section 2.5 for details). The model describes the probability of classifying the structural identity as a function of $\{\bm{\phi}(C_i),\bm{\phi}(C_j)\}$. Of the various metric learning methods proposed so far \cite{musgrave2020metric,wang2021deep}, we applied the Siamese network \cite{chopra2005learning} and KISS (\textit{keep it simple and straightforward!}) metric learning \cite{koestinger2012large}, a na\"{i}ve binary classifier, and a regression model that regresses the structure similarity value instead of $y_{ij}$. The binary classifier and the regressor were modeled as a conventional multi-layer perceptron (MLP) wherein the input variable is given by the absolute difference between two compositional descriptors, $|\bm{\phi}(C_i)-\bm{\phi}(C_j)|$. By comparing the generalization performance of the four metric learning methods mentioned above, we found that binary classification using the MLP outperformed others, as shown in Fig. S1. Therefore, hereafter, we report the results of the CSP using the binary classification neural network. 

\newpage
\subsection{Overall prediction scheme of the CSP method}
\begin{figure}[htp]
\centering
\includegraphics[width=12cm]{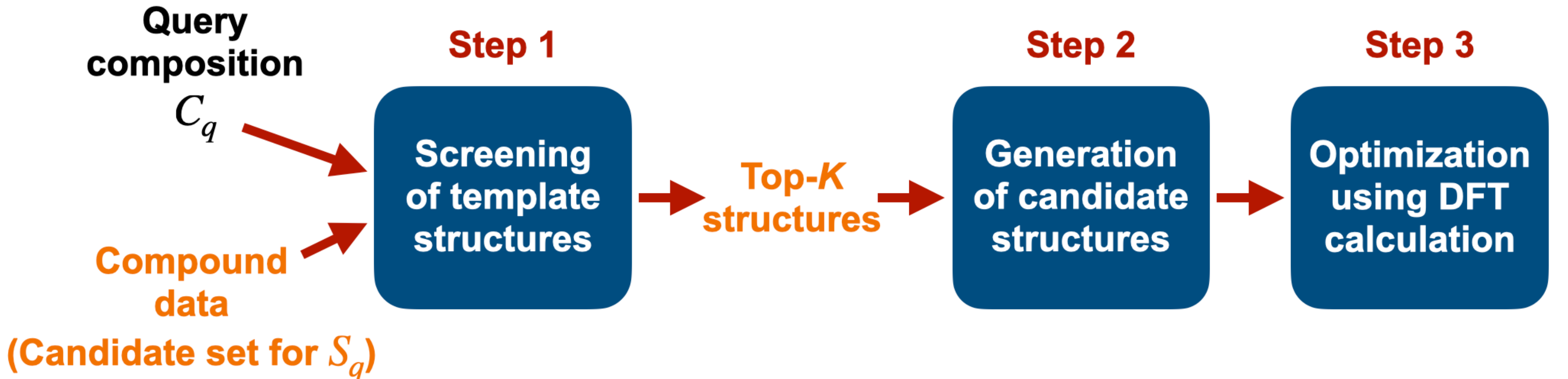}
\caption{Overall scheme of the CSP method.}
\label{fig:fig2}
\end{figure}

The overall scheme of the CSP method consists of three steps, as shown in Fig. \ref{fig:fig2}. The details of each step are described below.

\subsection*{Step 1: High-throughput screening of template crystal structures}
For the CSP of a query composition $C_q$, the screened candidates are limited to crystals with the same compositional ratio. For example, if \ce{Li3PS4} is given as the $C_q$, only crystal structures with a composition ratio of 3:1:4 (the order does not matter) are used as the candidate templates. In the applications shown below, the number of structures to be screened varies from 1 to 3,895 (see Fig. S6). The stable structures with the top-$K$ chemical compositions judged to have the highest output probability or structure similarity are selected as the template structure for the query composition.

\subsection*{Step 2: Generation of candidate crystal structures}
 A crystal structure is created by assigning the constituent elements of the query composition to the atomic coordinates of each selected template. Elements with the same composition ratio between the template and the query are assigned. It is important to note that, when one or more elements have the same composition ratio, the assignment is not uniquely determined. For example, there are two possible combinations in the assignment of \ce{A1B1} to \ce{C1D1} as \{A→C \& B→D\} and \{A→D \&
B→C\}. In such cases, we replaced a pair of elements with the most similar physicochemical properties. Specifically, the element similarity is defined as the Euclidean distance of the 19 elemental descriptors (Fig. S3). The crystal structure generated inherits the lattice parameters and atomic coordinates of the template structure. Note that this exchange rule does not always guarantee the best solution. Instead, all possible exchanges could be tested.

\subsection*{Step 3: Geometry optimization using the DFT calculation}
Finally, the $K$ candidate structures of $C_q$ are locally optimized by performing DFT calculations. DFT calculations were performed using the Vienna ab initio simulation package (VASP, version 6.1.2) \cite{kresse1996efficient}, combined with the projector augmented wave  pseudopotentials \cite{blochlProjectorAugmentedwaveMethod1994}. The exchange-correlation functional was considered with the generalized gradient approximation based on the Perdew-Burke-Ernzerhof method \cite{PerdewGeneralizedGradientApproximation1996}. The Brillouin zone integration for unit cells was automatically determined using the $\Gamma$-centered Monkhorst-Pack meshes function implemented in the VASP code. To generate the inputs of VASP calculations, the ``MPStaticSet'' and ``MPRelaxSet'' presets implemented in Pymatgen \cite{ong2013PythonMaterials} were used.

\subsection{Chemical composition descriptor}
We calculated the compositional descriptor, $\bm{\phi}(C_i)\in\mathbb{R}^d$, using XenonPy \cite{xenonpy_1,liu2021machine}. XenonPy is an open-source Python library for materials informatics that provides 58 physicochemical features for each element (Fig. S4). For each element-level feature, the compositional descriptor is calculated by taking summary statistics of constituent elements with the composition ratios such as the weighted mean, weighted sum, weighted variance, and min- and max-pooling. Thus,  for a given chemical composition, a 290-dimensional ($58\times5$) descriptor vector is defined.

\subsection{Preparation of structure similarity labels}
Metric learning relies on the supervision of the binary class label $y_{ij}$, which indicates whether a pair of crystal structures are similar or dissimilar. The class label is calculated as follows: (1) quantify the crystal structure similarities of all compound pairs, and (2) binarize the similarity measures by applying a prescribed threshold $\tau$.

To calculate the structure similarity, we encoded a given structure using the site fingerprint with local structure order parameters \cite{zimmermann2020local} (implemented in Matminer \cite{ward2018matminer}, an open-source toolkit for materials data mining). By evaluating the degree of resemblance of the coordination environment of an atomic site to the preset-coordination motifs, we obtained a vector-type descriptor (site fingerprint) for each atomic site in the crystal structure. Then, a crystal structure descriptor was calculated by taking the summary statistics of the site fingerprints across all atomic sites in the crystal structure. We used the mean, standard deviation, minimum, and maximum as the summary statistics. Finally, the structure similarity was calculated as the Euclidean distance between the crystal structure descriptors. The similarity measure uses only the topological features of the atomic coordinates and does not use any information about the elemental composition. Note that any similarity measure, such as the one proposed by Thomas et al. \cite{Thomas2021}, can be used for calculating the class labels. In this study, the structural similarity based on the site fingerprints was used because it is officially adopted by the Materials Project and is one of the standard similarity measures.

Following the procedure described above, we calculated 549,544,128 crystal structure dissimilarities between all pairs of the 33,153 stable compounds in the Materials Project database \cite{jain2013commentary,MaterialsProject} (version released on 11/21/2020). A histogram of dissimilarities is shown in Fig. S5. We considered an appropriate threshold for the binarization of similarity measures. The trade-off in the occurrence of false positives and false negatives should be considered when choosing the threshold value: a large threshold value increases the number of false negative cases where structurally similar structures are judged to be dissimilar; a small threshold increases the number of false positives.  If the threshold is too small, the number of positive instances (structurally similar pairs) becomes very small, making the treatment of imbalanced data difficult.

To determine the value of $\tau$ that appropriately balances the trade-off, we tested the binarization by varying $\tau \in \{ 0.01, 0.1, 0.2, 0.3, 0.4, 0.5, 0.6, 0.7\}$. For each $\tau$, we examined the number of instances classified as ``similar'' and the proportion of compounds that appeared at least once in the class ``similar'' (Table S1).  At $\tau = 0.3$, approximately 80\% of all the compounds appeared at least once in the class ``similar''. The remaining 20\% were judged to have no similar pairs. This means that the structures of these 20\% cannot be predicted using the substitution-based method. In contrast, stable structures of 80\% of the crystals can be determined using the substitution-based method. Based on these considerations, we set the threshold to $\tau = 0.3$.

\subsection{Experimental procedure}
From the 126,335 inorganic compounds in the Materials Project database, we obtained 33,153 stable compounds with an energy above the hull equal to zero. To benchmark the predictive performance of the proposed CSP, 38 crystals were selected, taking into account the diversity of space groups, structures, constituent elements, the number of atoms per unit cell, and their application domains; the number of atoms per unit cell was distributed in the range of 2 to 104. The crystal structure data (CIF files) are provided in the Supplementary Data (see Code and Data Availability).

The rest 33,115 compounds, which were not used for the benchmark, were randomly divided into 10,000 for training, 2,000 for validation, and 21,115 for testing in the process of metric learning. Of the 10,000 training compounds, (49,995,000 pairs, ${}_{10,000} \mathrm{C}_2$), 421,000 pairs were categorized as similar at $\tau = 0.3$. To eliminate the imbalance between the number of positive and negative instances in the training of the classifier, 421,000 negative instances were randomly selected from the 49,574,000 dissimilar groups. Following the same procedure, the validation and test sets were selected so that the number of positive and negative instances was equal, resulting in a total of 32,050 and 3,782,728 pairs, respectively.

The model input was defined as the absolute difference between the 290-dimensional compositional descriptors, $|\bm{\phi}(C_i)-\bm{\phi}(C_j)|$, and the output was given by the similarity label $y_{ij}$. The binary classifier was independently trained five times using randomly selected training and validation datasets. During each training, the hyperparameters were adjusted to provide the highest prediction accuracy for the validation set (Supplementary Information). The ensemble of these five models, $f_1, \ldots, f_5$, was used to produce the predicted class label. The class probability of being classified into similar pairs is given by $\hat{f}(|\bm{\phi}(C_i)-\bm{\phi}(C_j)|) = \frac{1}{5}\sum_{b=1}^{5} f_b(|\bm{\phi}(C_i)-\bm{\phi}(C_j)|)$.
For the set of candidate templates, we used all 33,115 stable structures except the 38 benchmark query compositions.

For a given query composition, according to the magnitude of the class probability of being classified into similar pairs in which the 33,115 candidate compounds with known crystal structures were screened out, we identified the top five template structures with a probability greater than 0.5. We then constructed candidate crystal structures as described above, which were optimized using the full structural relaxation in DFT.

In addition to the 38 benchmark compositions, 50 stable structures, randomly selected from 21,115 test samples, were selected as benchmarks. As candidate templates, 33,103 stable structures were used, excluding the 50 query compositions.

\section{Results}
\label{sec:section3}
\begin{figure}[H]
\centering
\includegraphics[width=6cm]{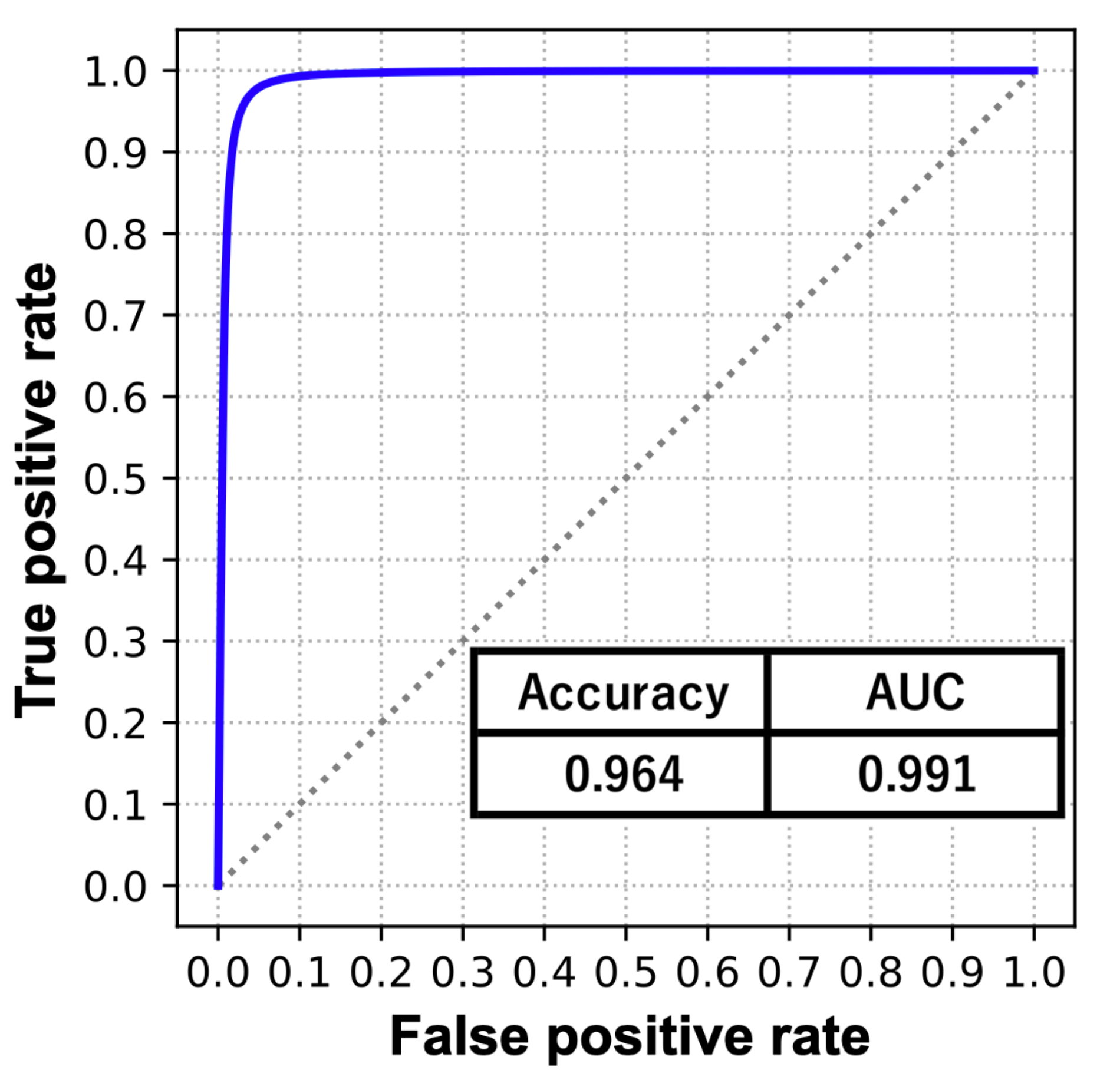}
\caption{ROC curve and performance metrics (accuracy and AUC) in the prediction of the structural identity  using the fully connected neural networks. 
}
\label{fig:fig3}
\end{figure}

The performance of the ensemble prediction that used five different neural networks was measured based on  the dataset consisting of the similar and dissimilar pairs of the test 21,115 compounds. The receiver operator characteristic (ROC) curve \cite{hanley1982meaning} according to the varying thresholds of the classification probability is shown in Fig. \ref{fig:fig3}. The area under the curve (AUC) \cite{hanley1982meaning} and the prediction accuracy were 0.991 and 96.4\%, respectively. The sensitivity (recall or true positive rate) and specificity (true negative rate) were 96.3\% and 96.6\%, respectively.

Here, bias in the test dataset could bring bias in the observed predictive performance. Fig. S7 showed a histogram of the top-50 most frequent composition ratios for the 33,153 compounds containing a total of 2,016 different composition ratios. It is observed that the dataset is highly biased to certain composition ratios. In order to examine the effect of bias in composition ratios, we evaluated the performance of the trained models using a different test dataset that balanced sample sizes for different composition ratios. To be specific, we randomly removed test samples with an over-represented composition ratio containing more than 50 samples such that the resulting sample size became equal or less than 50. The AUC and the prediction accuracy for this test set were 0.995 and 94.6\%, respectively. To conclude, the bias in the composition ratio has almost no effect on the accuracy of the prediction.

According to the performance tests shown above, the similarity of the stable structures of the  two given chemical compositions can be predicted with a considerably high accuracy. We applied this similarity prediction model to identify the known stable structures of the 38 benchmark crystals. The results are summarized in Table \ref{table1}. The proposed method was applied to select a maximum of five template structures, which were then subjected to element substitution to produce a set of candidate crystal structures. Hereafter, this set is denoted as $\mathcal{C}_{5}$. As mentioned above, only the structures with the same composition ratio as the query among the total 33,115 stable structures in the Materials Project were considered as candidates during the screening. We denote this set as $\mathcal{C}$. The number of candidate structures selected here varies greatly depending on the query compositions: for all the stable structures in the Materials Project, the mean and median numbers of structures with the same composition ratio were 322.0 and 952.0, respectively, and the maximum and minimum numbers were 3,895 and 0, respectively. By definition, a crystal structure with the absence of matching structures is unpredictable (1,051/33,153) by the proposed method. The distribution of the number of matches is shown in Fig. S6. It can be seen that the number of candidate structures is considerably reduced after composition ratio matching.

The CSP method could not propose any templates for three out of 38 query compositions, \ce{NaCaAlPHO5F2}, \ce{MgB7}, and \ce{Ba2CaSi4(BO7)2}: none of the candidates had the same composition ratio as \ce{NaCaAlPHO5F2} in the 33,115 candidates; for \ce{MgB7} and \ce{Ba2CaSi4(BO7)2}, none of the candidates had class probabilities greater than 0.5. Table \ref{table1} summarizes the prediction results for the remaining 35 query compositions. We investigated the dissimilarity value of the closest structure in $\mathcal{C}_{5}$ to the known stable structure (the third column), which was compared with the minimum dissimilarity value of the best template among all candidates in $\mathcal{C}$ (the second column). The structural dissimilarities were calculated with the local structure order parameter presented in \cite{zimmermann2020local}. This measure was defined only on atomic coordinates, ignoring the information of element types. For quantitative evaluation, the rank of the minimum dissimilarity value of the top five candidates was calculated with respect to the dissimilarities of $\mathcal{C}$ (the fourth column). The model could select the best template, which is the closest to the true stable structure in $\mathcal{C}$, with an accuracy of approximately 51.4\% (=18/35) by screening out the top five candidates. 
Only two cases, \ce{Li_3 PS_4} and \ce{BN}, had ranks higher than 30. With the top five candidates, the proposed method succeeded in identifying templates that are almost equivalent to the best template.

\begin{table}[H]
\centering

\caption{Results of the CSP for the 38 benchmark systems. The first column lists the 35 query compositions which were predictable. The second column lists the minimum dissimilarities of all candidate structures that have the same composition ratio as the query. The third column presents the minimum structural dissimilarities for the top five identified structures with respect to the true stable structure. The fourth column lists the rank of the minimum dissimilarity of the top five candidates in all candidates. For example, the top five candidates of \ce{Al2O3} are ranked 2nd out of 297 candidates; so the cell is marked 2/297.
It is indicated in the fifth column whether the true structure is included in the top five predicted structures relaxed by DFT ($\checkmark$ and $-$ indicate success and failure, respectively).
}
\label{table1}

\scalebox{0.6}{
\begin{tabular}{r|cccc}
        \toprule
        \textbf{Composition} & \textbf{Min. dissimilarity of all candidates}  & \textbf{Min. dissimilarity of top 5}  & \textbf{Rank} & \textbf{Prediction success} \\
        \midrule
        \ce{Ag8GeS6} &  0.214  & 0.214 & 1/34 & $-$ \\
        \ce{Al2O3} &  0.067  & 0.093 & 2/297 & $\checkmark$ \\
        \ce{BN} &  1.726  & 3.292 & 683/960 & $-$ \\
        \ce{Ba(FeAs)2} &  0.091  & 0.176 & 9/1424 & $\checkmark$ \\
        \ce{Bi2Te3} &  0.293  & 0.293 & 1/297 & $\checkmark$ \\
        \ce{C} &  1.769  & 1.975 & 3/87 & $-$ \\
        \ce{Ca14MnSb11} &  0.083  & 0.096 & 2/13 & $\checkmark$ \\
        \ce{CaCO3} &  0.054  & 0.077 & 3/1000 & $\checkmark$ \\
        \ce{Cd3As2} &  0.19  & 0.19 & 1/297 & $\checkmark$ \\
        \ce{CoSb3} &  0.068  & 0.068 & 1/1042 & $\checkmark$ \\
        \ce{CsPbl3} &  0.129  & 0.129 & 1/1000 & $\checkmark$ \\
        \ce{Cu12Sb4S13} &  0.24  & 0.24 & 1/1 & $\checkmark$ \\
        \ce{Fe3O4} &  0.216  & 0.216 & 1/152 & $-$ \\
        \ce{GaAs} &  0  & 0 & 1/960 & $\checkmark$ \\
        \ce{GeH4} &  0.383  & 0.639 & 22/171 & $-$ \\
        \ce{La2CuO4} &  0.022  & 0.022 & 1/821 & $\checkmark$ \\
        \ce{Li3PS4} &  0.851  & 1.216 & 33/250 & $-$ \\
        \ce{Li4Ti5O12} &  0.282  & 0.282 & 1/8 & $-$ \\
        \ce{LiBF4} &  0.302  & 0.592 & 6/983 & $-$ \\
        \ce{LiCoO2} &  0.199  & 0.207 & 5/3895 & $-$ \\
        \ce{LiFePO4} &  0.113  & 0.13 & 2/327 & $\checkmark$ \\
        \ce{LiPF6} &  0.046  & 0.297 & 6/242 & $\checkmark$ \\
        \ce{Mn(FeO2)2} &  0.022  & 0.022 & 1/821 & $\checkmark$ \\
        \ce{Si} &  0  & 2.304 & 7/87 & $-$ \\
        \ce{Si3N4} &  0.269  & 0.269 & 1/152 & $-$ \\
        \ce{SiO2} &  0.167  & 0.167 & 1/1151 & $-$ \\
        \ce{SrTiO3} &  0.395  & 0.643 & 16/1000 & $\checkmark$ \\
        \ce{TiO2} &  1.015  & 1.401 & 20/1151 & $-$ \\
        \ce{V2O5} &  0.753  & 1.865 & 41/85 & $-$ \\
        \ce{VO2} &  0.077  & 0.077 & 1/1151 & $\checkmark$ \\
        \ce{Y3Al5O12} &  0.014  & 0.014 & 1/49 & $\checkmark$ \\
        \ce{ZnO} &  0.006  & 0.062 & 5/960 & $\checkmark$ \\
        \ce{ZnSb} &  0.316  & 0.316 & 1/960 & $\checkmark$ \\
        \ce{ZrO2} &  0.131  & 0.131 & 1/1151 & $\checkmark$ \\
        \ce{ZrTe5} &  0.039  & 0.039 & 1/132 & $\checkmark$ \\
        \bottomrule
\end{tabular}
}

\end{table}

\begin{figure}[H]
\centering
\includegraphics[width=1.0\linewidth]{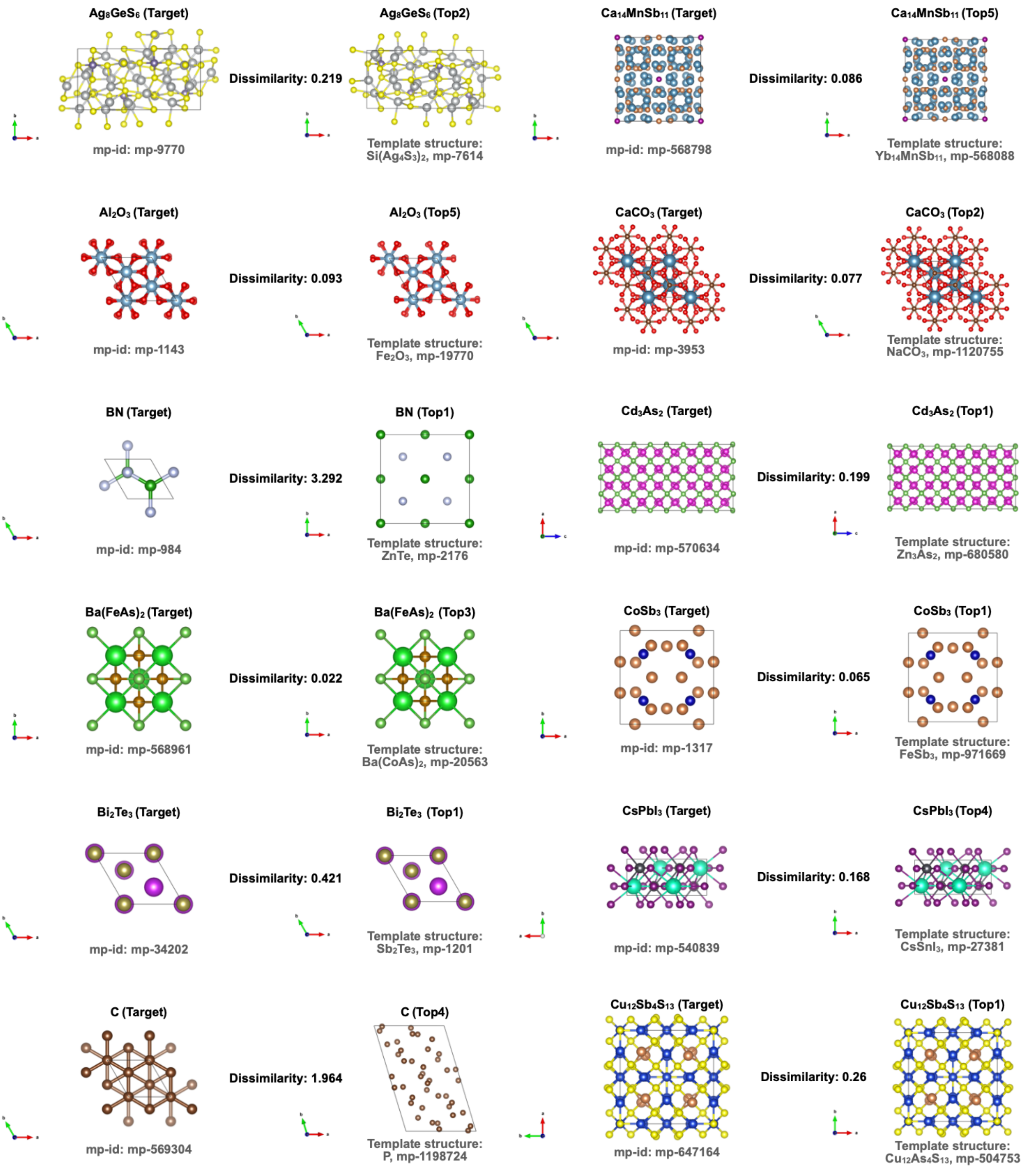}
\caption{12 examples of the predicted and true crystal structures. The closest predicted structure to the true structure among the top five candidates (depicted with VESTA \cite{momma2011vesta}) is shown. For all the results of the 35 test cases, see Fig. S8. 
}
\label{fig:fig4}
\end{figure}

After the element substitution, the candidate crystal structures were locally optimized using DFT calculations. The predicted and true crystal structures for 12 arbitrarily selected queries are shown in Fig. \ref{fig:fig4}. Of the top five, the predicted structure closest to the true structure is illustrated. The predicted crystal structures for all the 35 queries are shown in Fig. S8. It can be seen that highly complex crystal structures consisting of a large number of atoms per unit cell could be successfully predicted, thereby demonstrating a definitive improvement over ordinary CSP programs. 
We determined whether the true structure was included in the top five relaxed predicted structures by performing a visual inspection. As summarized in Table \ref{table1} (the fifth column), in 21 out of the 35 queries (60\%), the top five predicted structures contained the true stable structure.

We performed the same analysis for the 50 randomly selected benchmark compositions. The CSP method could not propose any templates for 4 out of the 50 query compositions, \ce{Ca2As4Xe5F34}, \ce{CsAg2(B5O8)3}, \ce{K4U5Te2O21}, and \ce{Mn3Sb5(IO3)3}, since none of the 33,103 candidates contained samples with the same composition ratio as the 4 query compositions. The prediction results for the remaining 46 query compositions are summarized in Table S2. As in the fourth column of Table S2, the top five candidates proposed by the model contained the best templates for approximately 76.1\% (=35/46) of the query compositions. Furthermore, as summarized in the fifth column of Table S2, the structural relaxation of the top five candidates using DFT succeeded in reconstructing 40 stable structures out of the 46 queries (87.0\%). The predicted crystal structures for the 46 queries are shown in Fig. S9.

Based on the results for the 81 (35+46) compounds in the two query sets, we examined the key factors that determine success or failure. First, for the failed and successful query compositions, the median, mean, and standard deviation of the number of elements were calculated. The respective values of the statistics were 3.0, 2.92, 0.63 (for success) and 2.0, 2.45, 0.80 (for failure). Thus, there was no significant difference between the successful and unsuccessful cases. For the number of atomic sites per unit cell, the three statistics also showed no significant difference: 24.0, 40.1, 47.5 (for successful queries), and 20.0, 35.3, 39.7 (for unsuccessful queries).

In contrast, when the dissimilarity between the best template among the top five template structures and the true structure (the third column in Table 1 and Table S2) was compared between success and failure, the median, mean, and standard deviation showed remarkable differences: 0.073, 0.142, 0.176 (for success) and 0.591, 0.982, 0.918 (for failure). Therefore, in the proposed CSP, the selection of a template structure sufficiently close to the true structure is the dominant factor that determines the success or failure.

Finally, we estimated the extent to which elemental substitution could cover the entire crystal system in predicting stable crystal structures. Although only 88 query compositions were tested, we investigated the ability to identify the best templates for the 21,115 stable crystals selected randomly from the Materials Project. Each of the 21,115 crystals was used as a query to select a candidate template from the known stable structures . Here, we predicted the stable structure of the query composition without performing DFT structural relaxation (Table \ref{table2}). For the top 50 identified templates, 10,829 out of 21,115 (=51.3\%) stable crystals were found to have the true stable structures within a radius of 0.1 of the structural dissimilarity. This corresponds to a prediction accuracy of 99.2\% ($=10,829/10,914$). Table \ref{table2} also summarizes the performance of detecting the best templates with the top $K$ predicted templates when the number of identified templates ($K$) was varied from 1 to 50, and the radius threshold was varied from 0.1 to 0.3. As shown in Table \ref{table1} and Table S2, when a template structure with a structural dissimilarity less than 0.1 could be selected, the proposed method could identify the true stable structures with 100\% accuracy ($=36/36$) by performing structural relaxation with DFT. Therefore, we estimate that approximately 51.3\% ($51.3 \times 1.0$) of the entire crystal system can be predicted using the proposed method. Moreover, if, for example, the threshold of the radius was set to 0.2, the proportion of the best templates falling within the given radius of the top 50 candidates was 66.8\%, and the structural relaxation using DFT converged to the true stable structures for approximately 97.9\% ($=47 /48$) of the crystals . Therefore, it is estimated that approximately 65.4\% ($66.8 \times 0.979$) of the entire crystal system can be predicted by our method.

\begin{table}[H]
\centering

\caption{Results of the best template prediction for 21,115 unique crystal systems in the Materials Project. The first column lists the threshold $\tau$ for the dissimilarities. The second column denotes the proportions of systems that had at least one candidate with a dissimilarity of less than $\tau$ in all candidates. The rest of the columns denote the proportions of the systems that had at least one candidate with a dissimilarity of less than $\tau$ in the top 1, 5, 10, 20, 30, and 50 suggested candidates, respectively.}
\label{table2}

\scalebox{0.8}{
\begin{tabular}{r|ccccccc}
        \toprule
        \textbf{$\tau$} & \textbf{All candidates}  & \textbf{Top 1}  & \textbf{Top 5} & \textbf{Top 10} & \textbf{Top 20} 
        & \textbf{Top 30} & \textbf{Top 50} \\
        
        \midrule
        0.1 &  51.7\%  & 31.7\% & 44.8\% & 47.7\% & 49.7\%
        & 50.5\% & 51.3\% \\
        0.2 &  68.1\%  & 46.0\% & 60.7\% & 63.5\% & 65.3\%
        & 66.0\% & 66.8\% \\
        0.3 &  76.4\%  & 55.5\% & 69.8\% & 72.4\% & 73.9\%
        & 74.5\% & 75.1\% \\
        \bottomrule
\end{tabular}
}

\end{table}

\section{Conclusion }
\label{sec:section4}
We proposed a CSP method based on metric learning of crystal structure similarity. The prediction is made by selecting crystal structures, which are predicted to be similar to the stable structure of a given query composition, from the existing crystal structures in the database. In materials science, most crystals have been discovered by element substitution of previously discovered crystals. The proposed method can be considered as a machine learning alternative to traditional protocols in the discovery of new materials. Compared to existing methods, the most significant difference is that the proposed method does not involve any first-principles calculations, except in the final step of locally optimizing the proposed structure. Thus, the computational cost of the proposed method is significantly lower than existing methods.

Finally, we summarize the extensibility and limitations of the proposed method. Although we have focused on the prediction of stable structures, the current method may also be applicable to the prediction of metastable structures. In principle, the present method can be used to predict the identity of metastable structures in the same framework if the training instances including the metastable structures are created. The method relies on element substitution; therefore, it cannot be applied unless there is a template available for substitution. For example, as reported in this paper, the crystal structures of approximately 3.2\% (1,051/33,153) in the Materials Project have no template with the same composition ratio. Nevertheless, the present method is highly capable of identifying the template closest to the true structure present in a crystal structure database. Furthermore, as discussed in the Results section, at least 50-65\% of all crystal systems, including unique crystals without template structures, can be predicted using the substitution-based CSP. If the crystal structure database expands monotonically in the future, the application range of the substitution-based CSP method will also expand.

\section*{Code and data availability}
The Python code for crystal structure prediction is available on GitHub \cite{CSPML}. The code can be readily used for the general purpose of CSP. Furthermore, readers can reproduce all the results of the crystal structure prediction presented in this paper with the dataset contained in the repository. The crystal structure data (CIF files) for the true and predicted structures of the 38 benchmark sets and the 50 randomly selected benchmark sets are provided in the Supplementary Data.

\section*{Acknowledgments }
This work was supported in part by a MEXT KAKENHI Grant-in-Aid for Scientific Research on Innovative Areas (Grant Number 19H05820), a JSPS Grant-in-Aid for Scientific Research (A) 19H01132 from the Japan Society for the Promotion of Science (JSPS), and JST CREST Grant Number JPMJCR19I3. The authors are grateful to Hiromasa Tamaki, Tomoyasu Yokoyama,  Kensuke Wakasugi, Koki Ueno, and Satoshi Yotsuhashi from Panasonic Corporation for helpful discussions and generously providing us with a list of benchmark crystals.

\section*{Author Contributions}
Minoru Kusaba and Ryo Yoshida designed the research; Minoru Kusaba and Ryo Yoshida wrote the manuscript; Minoru Kusaba wrote the program and performed the analysis; Chang Liu performed DFT calculations and supervised the calculation results; Ryo Yoshida supervised the research. 

\section*{Additional Information}
Competing Interests: The authors declare no competing interests.

\bibliographystyle{elsarticle-num} 
\bibliography{cas-refs}





\end{document}